\documentclass[12pt,preprint]{aastex}
\newcommand{\etal}{et al. }
\newcommand{\teff}{{\mathrm{T}}_{\mathrm{eff}}}
\newcommand{\zzc}{ZZ~Ceti }
\newcommand{\msun}{M_{\odot}}

\begin{document}

\title{Ensemble Characteristics of the \zzc stars}

\author{Anjum S. Mukadam\footnote{Hubble Fellow}}
\affil{Department of Astronomy, University of Washington, Seattle, WA\,-\,98195-1580, U.S.A.; anjum@astro.washington.edu}
\author{M. H. Montgomery, D. E. Winget}
\affil{Department of Astronomy, University of Texas at Austin, Austin, TX -78712, U.S.A.}
\author{S. O. Kepler}
\affil{Instituto de F\'{\i}sica, Universidade Federal do Rio Grande do Sul, 91501-970 Porto Alegre, RS - Brazil.}
\author{J. C. Clemens}
\affil{Department of Physics and Astronomy, University of North Carolina, Chapel Hill, NC 27599, U.S.A.}

\begin{abstract}

We present the observed pulsation spectra of all known non-interacting \zzc stars (hydrogen atmosphere
white dwarf variables; DAVs) and examine changes in their pulsation properties across the instability strip. 
We confirm the well established 
trend of increasing pulsation period
with decreasing effective temperature across the \zzc instability strip.
We do not find a dramatic order of magnitude increase in the number of
observed independent modes in \zzc stars, traversing from the hot to the cool edge of the
instability strip; we find that the cool DAVs have one more mode on average compared to the hot DAVs.
We confirm the initial increase in pulsation amplitude at the blue edge, and
find strong evidence of a decline in amplitude prior to the red edge.
We present the first observational evidence that \zzc stars lose pulsation energy just before
pulsations shut down at the empirical red edge of the instability strip.

\end{abstract}


\section{Introduction}
Asteroseismology is the only systematic technique to study the insides of a star.
Pulsators are fortunately found all over the H-R diagram, and allow us the
opportunity to look inside different stars in various evolutionary stages.
White dwarf stars are the stellar remains of 98--99\% of stars in the sky \citep{Weidemann90},
and contain an archival record of their main sequence lifetime. Pulsating white dwarf stars
serve as effective instruments to harness this archival record. 

White dwarf spectra reveal that 80\% of them have atmospheres
dominated by hydrogen (DAs; \citealt{Fleminget86}). Hydrogen atmosphere DA white dwarfs
are observed to pulsate in an instability strip located in the temperature range
10800--12500\,K for $\log~g\approx 8$ \citep{Bergeronet95,Bergeronet04,KoesteraAllard00,KoesteraHolberg01,Mukadamet04b}.
Recent studies determine that the DA instability strip is only 
about 1000--1200\,K in width \citep{Mukadamet04b,Gianninaset05}.
These DA variables (DAVs)
are known as the \zzc stars, after the proto type \zzc (R\,548) of the class.

In this paper we gather the pulsation spectra of all known non-interacting \zzc stars with the purpose of
studying their ensemble pulsation characteristics, and how these properties change across 
the instability strip. Such a systematic study 
of the \zzc pulsators was first undertaken by pioneers in the field
such as \citet{Robinson80}, \citet{McGrawet81}, and \citet{WingetaFontaine82}. \citet{Clemens93}
was the first to demonstrate the distinct behavior of pulsation periods and
amplitudes as a function of temperature systematically for a significant sample of DAVs.
The sample size of known DAVs is now
three times larger than the sample used in the last such characterization of the \zzc 
pulsators by \citet{Kanaanet02}.

Also, we now have better and internally consistent temperatures for these pulsators,
calculated using the ML2/$\alpha=0.6$ convection 
description \citep[see][]{Bergeronet04,Kleinmanet04}. This prescription for convection gives
the best internal consistency between optical and ultra-violet (UV) effective temperatures, trigonometric 
parallaxes, V magnitudes, and gravitational redshifts \citep{Bergeronet95,KoesteraVauclair97,BergeronaLamontagne03,Fontaineet03}.
These two reasons compel us to
re-examine the pulsation characteristics of the instability strip.

\subsection{Defining our \zzc samples}
Of the 35 new \zzc variables published in \citet{Mukadamet04a}, HS0951+1312, HS0952+1816,
WD1443+0134, WD1524-0030, and WD2350-0054 do not have reliable spectroscopic $\teff $ and
$\log~g$ fits (see \citealt{Mukadamet04a, Mukadamet04b}).
Our sample of new \zzc variables discovered using spectra from the Sloan Digital Sky Survey (SDSS) then
consists of 30 DAVs from \citet{Mukadamet04a} and 11 DAVs from \citet{Mullallyet05}. For these 41 DAVs,
D. Eisenstein derived a homogeneous and internally consistent set of $\teff $ and $\log~g$
values from the SDSS using model
atmospheres from D. Koester \citep[see][]{Kleinmanet04}. We carefully and consistently re-analyzed all of our
time-series photometry data on these pulsators, acquired using the prime focus CCD camera 
Argos \citep{NatheraMukadam04}
on the 2.1\,m telescope at
McDonald Observatory.
We shall
henceforth refer to this set of 41 DAVs as the SDSS \zzc sample with homogeneous spectroscopic fits and
homogeneous time-series photometry.

\citet{Bergeronet04} and \citet{Gianninaset05} have published
internally consistent temperatures and $\log~g$ fits for 39 DAVs
using their latest model atmospheres; 
we will henceforth refer to this second set
of 39 DAVs as the BG04 \zzc sample.
We compiled a corresponding set of 39 pulsation spectra from the
literature, and via private communication with our colleagues. The seismic data of the BG04 \zzc sample
was acquired by different observers using different instruments and telescopes. However a substantial amount of
time-series data exists on most DAVs in this sample, and we utilized practically
all published pulsation spectra to carefully derive well-averaged values of weighted mean period 
and pulsation amplitudes which we present in this paper.

\citet{Mukadamet04b} show evidence of a relative shift of about 200\,K
between the SDSS and BG04 \zzc instability strips, which also differ in shape and width.
The spectra of \zzc stars in these
samples were analyzed using different techniques and with different 
model atmospheres. As homogeneity is imperative, we cannot
merge these two samples for analyses that involve the spectroscopic temperature.
However we can merge these two samples when formulating plots
based solely on seismic data such as Figure 3.

	During the writing of this manuscript, 25 new \zzc stars were submitted for publication
in separate papers: Silvotti \etal 2005a, Castanheira \etal 2005, Kepler \etal 2005b, Gianninas \etal 2005,
and Silvotti \etal 2005b.
This brings the total number of non-interacting \zzc stars known to 107.
Twenty of the new DAVs in these papers come from the SDSS. Their spectroscopic $\teff$ and
$\log~g$ values were derived using the same technique as the 41 SDSS DAVs discovered previously,
but with a different algorithm (version auto23 vs. auto21 used earlier). 
Hence we cannot include the new DAVs in our SDSS
\zzc sample for plots based on effective temperature, such as Figures 1 and 2.
Also, we find that the
main purpose of the pulsation spectra presented in these papers is to demonstrate discovery of
variability. Such pulsation spectra may be incomplete and are not well suited for direct inclusion in most plots of 
this paper where ensemble homogeneity
plays a big role\footnote{We added three new DAVs HS1039+4112 \citep{Silvottiet05a}, 
G232-38 \citep{Gianninaset05}, and GD133 \citep{Silvottiet05b}
to the original BG04 sample of 36 stars; this does not significantly affect the homogeneity of the BG04 sample as the new DAVs constitute less than 10\%
of the sample. However adding 20 new SDSS DAVs to the SDSS \zzc sample of 41 stars would significantly affect
the homogeneity of the sample.}.
However we will include these variables in Figure 3 (based on seismic data) for completeness, and show the net 
effect of the inclusion on the original plot.

\subsection{Methodology: using only linearly independent pulsation frequencies}
\citet{Brickhill92}, \citet{Brassardet95}, \citet{Wu01}, and \citet{Montgomery05}
show that the non-linear pulse shapes observed in some variable white dwarfs may
arise as a result of relatively thick convection zones.
Many frequencies are evident in these stars because the
non-linearities appear as harmonics and linear combinations in our Fourier transforms.
We are interested in studying the pulsation characteristics of self-excited 
real modes in the context of this paper. Hence we will disregard all harmonics
and linear combination frequencies in the observed pulsation spectra throughout this paper.

When our program detects a linear combination or harmonic in an observed pulsation spectrum,
we typically
hand-pick the lowest amplitude mode as the linear combination mode.
However if we find a mode
involved in two or more linear combinations, we consider it to be a linear combination mode even if it is not
the lowest amplitude mode. Our simplistic approach
could lead us to incorrect choices; we could be misinterpreting resonances as
linear combination modes in a few cases. However any small discrepancies in values
evaluated for a single star should have a minimal impact on the conclusions we draw from the ensemble.

\section{Changes in pulsation properties across the \zzc instability strip}
Short pulsation periods (100--300\,s) in the \zzc stars typically have pulsation amplitudes 
of a few percent or smaller. Long pulsation periods (600--1000\,s) typically have large
amplitudes that can be as high as 10\%.
This correlation between the pulsation periods and
amplitudes of the \zzc stars has been established for a long time \citep{Robinson79,Robinson80,McGraw80,Fontaineet82}.
However the correlation of these properties with temperature
had to wait more than a decade for sufficiently accurate determinations of \zzc temperatures.
Subsequently the distinct behavior of pulsation periods and
amplitudes as a function of temperature was systematically 
demonstrated for a significant sample of DAVs by \citet{Clemens93}
and more recently by \citet{Kanaanet02}.
We now demonstrate and discuss these trends for the new as well as previously known
DAVs of our two homogeneous samples. 
We will begin with showing how the number of observed independent modes changes
across the instability strip.

\subsection{Number of observed independent modes}
We expect a larger (than intrinsic) scatter in the number of
independent modes we measure for an ensemble of \zzc stars due to varied detection efficiency. This partially arises
because we use different instruments on different telescopes, or observe for different durations of time,
and also due to the variety of weather conditions and extinction values. 
A fraction of the scatter in our detection efficiency is caused
because \zzc stars have different magnitudes and different pulsation amplitudes, and the pulsation amplitudes
depend on the effective wavelength of observation \citep{Robinsonet82}. 
                                                                                               
        By using the SDSS \zzc sample alone, we reduce some of this expected scatter (see subsection 1.1). 
To reduce the scatter additionally,
we choose to limit ourselves to the magnitude range
$17.8\,\leq\,g\,\leq\,18.8$ and to a stellar mass $\msun\,<\,1.0\,\msun$ to exclude
intrinsic low amplitude DAVs (see section 2.3). Reducing 
our magnitude range any more will take us further into
the realm of small number statistics. For the subset of 23 SDSS DAVs that fall in the chosen range, we find
an average of 2.5\,modes/star for the hotter \zzc stars (11300--11800\,K), and
a slightly larger value of 3.1\,modes/star for the cooler \zzc stars (10800--11300\,K). 
If we consider 
all the SDSS DAVs in our sample, excluding the three massive variables (WD0923+0120, WD1711+6541, \& WD2159+1322),
then we find an average of 
2.9\,modes/star for the hotter \zzc stars (11300--11800\,K) and 3.3\,modes/star for the 
cooler \zzc stars (10800--11300\,K) up to
an amplitude limit of order a few mma. This may seem
contrary to published literature, but
we are attempting to restrict ourselves to independent self-excited modes only, rather than plot the
total number of observed periodicities in the star.

The period spacing for
nonradial g-modes with $k\leq$\,3 is larger than the asymptotic limit.
This increase in the density of available eigenmodes as a function of period
is consistent with the increase we find in the number of observed independent modes,
comparing the hotter \zzc stars to the cooler \zzc stars.

\subsection{Observed pulsation periods}
Pulsations in \zzc models (and in stars in general) are self-driven oscillations. 
Stochastic noise frequencies coincident
with the eigenfrequencies are amplified by the driving mechanism
to observable amplitudes. The blue edge of the \zzc instability strip 
occurs when the star is cool enough to have a hydrogen partial
ionization zone sufficiently deep to excite global pulsations.

Our models suggest that the driving frequency is governed
by the thermal timescale at the base of the hydrogen partial 
ionization zone (see \citealt{Cox80,Unnoet89}).
As the model star cools, the base of the partial
ionization zone moves deeper in the model, and the thermal timescale
increases \citep{Winget82,Hansenet85}. Numerical calculations by Montgomery (2005) 
show that the thermal timescale at the base of the ionization zone $\tau_{th}$ is proportional
to $\teff^{-90}$. However
recent investigations by \citet{Kimet05} suggest that the thermal timescale
is not consistent with the mean pulsation period, although both quantities increase in the models
as we traverse from the blue to the red edge of the instability strip.

Figure 1 shows the mean pulsation period as a function of effective temperature for DAVs in both the
SDSS and BG04 \zzc samples. We determine the Weighted Mean Period (WMP) for each DAV by weighting each 
period with the corresponding observed amplitude (see equation 1).

\begin{equation}
\mathrm{Weighted\,Mean\,Period\,(WMP)} = \frac{\sum_i P_i A_i}{\sum_i A_i}
\end{equation}
\clearpage
\begin{figure}[!ht]
\figurenum{1}
\epsscale{1.0}
\plotone{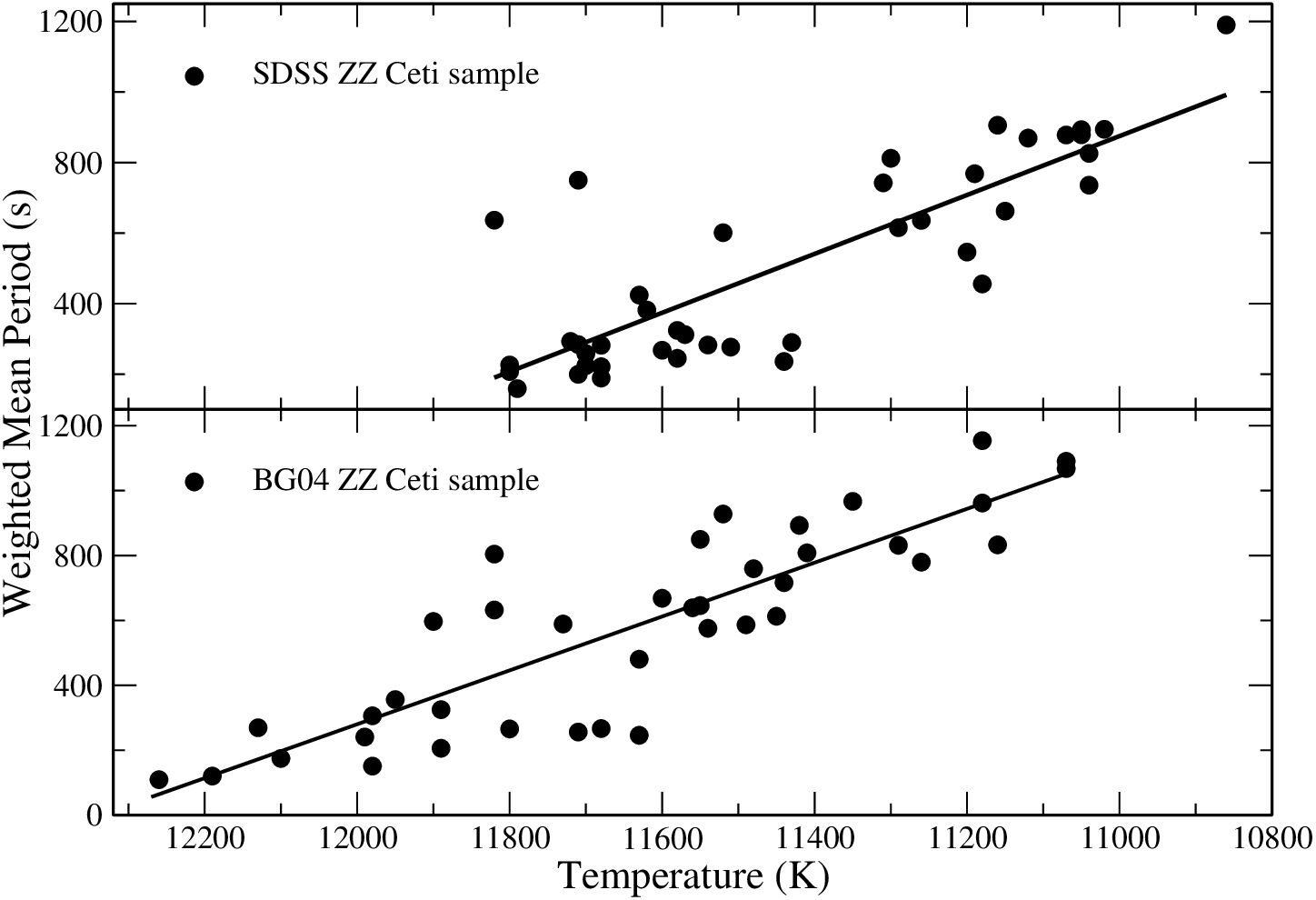}
\caption{We show the weighted mean period of 41 newly discovered \zzc stars from the SDSS (top)
and 39 \zzc stars from the BG04 sample (bottom) vs their spectroscopic temperature; each period 
was weighted by the observed amplitude. We note the distinct increase in mean pulsation
period as DAVs cool across the \zzc instability strip.}
\end{figure}
\clearpage

We can now appreciate the significance of Figure 1; the weighted mean pulsation period
provides us with a method of determining
the temperature of a \zzc star independent of spectroscopy.
The respective equations of the best fit lines through the SDSS and the BG04 \zzc samples
shown in Figure 1 are: \newline
${\mathrm{WMP}}_{\mathrm{SDSS}}$ = -0.835$^{\pm 0.089}$ $\times {\teff}_{\mathrm{SDSS}}$ + 10060$^{\pm 1020}$\newline
${\mathrm{WMP}}_{\mathrm{BG04}}$ = -0.830$^{\pm 0.079}$ $\times {\teff}_{\mathrm{BG04}}$ + 10240$^{\pm 920}$
                                                                                              
where WMP is in units of seconds and $\teff$ is in units of Kelvin. It is reassuring to see an agreement in the values of the slope for both the SDSS and BG04 \zzc samples, and
gives us more confidence in using this period-temperature correlation as a
means to measure the effective temperature of a DAV. We can now determine
a temperature for the five variables with unreliable spectroscopic fits 
we excluded earlier, HS0951+1312, HS0952+1816,
WD1443+0134, WD1524-0030, and WD2350-0054 (see Table 1 in the appendix).

We compute a sum of the squares of the {\bf horizontal} offsets of the points in Figure 1
respectively from the above best fit lines, and arrive at
a seismic estimate of the average uncertainty in our spectroscopic temperature.
We determine ${\sigma}_{\teff}$\,=\,188.3\,K
for the SDSS \zzc sample, and ${\sigma}_{\teff}$\,=\,185.4\,K for the BG04 \zzc sample.
It is not surprising that the average $\teff$ uncertainties in both these samples turned out to be similar
because \zzc stars in both samples define an instability strip of the same width $\sim$1000--1200\,K\citep{Mukadamet04b,Gianninaset05}.
We can expect that if the average $\teff$ uncertainty in the SDSS sample had been significantly larger than that
of the BG04 sample, the empirical width of the SDSS instability strip would also have been 
correspondingly larger.

Our current temperature determinations are getting better and may be
internally consistent to about 200\,K at present. Our
empirical measure of the width of the instability strip appears to have reduced
from 1500\,K \citep{Bergeronet95,KoesteraAllard00} to about 1000--1200\,K \citep{Mukadamet04b,Gianninaset05}.
But even with our current atmospheric models
we do find exceptions to this period-$\teff$ correlation, where the spectroscopic temperature
does not match the pulsation properties
of the \zzc star. For example, \citet{Mukadamet04a,Mukadamet04b} report the \zzc star
WD2350-0054 with a dominant period close to 300\,s as an unusual pulsator because its SDSS spectroscopic
determination places it below the red edge of the instability strip at $\teff = 10350 \pm 60$\,K.
\citet{Bergeronet04} report a known complex pulsator G29-38 \citep{Kleinmanet98} as a
hot \zzc star with $\teff=11820$\,K.
Better S/N spectra of these and other such \zzc stars may reveal temperatures consistent with
their observed periods.
Unless the pulsation characteristics themselves reveal curious features, it seems premature to conclude
that the \zzc star in question (with a spectroscopic temperature inconsistent with its pulsation spectrum)
is strange or abnormal in any way.

\subsubsection{Weighted Mean Period as a temperature indicator}
Measuring the pulsation periods of a \zzc star is a model-independent 
and straight forward process\footnote{While it is true that aliases complicate the
determination of pulsation periods from single site data, this ambiguity may change the
value of the period typically by only a few seconds.}
as opposed to a spectroscopic determination of its temperature. 
The instability strip spans 1000--1200\,K in
temperature and about 1300\,s in period.
The internal uncertainty in measuring temperature using spectra is typically about 200\,K,
which is 17--20\% of the width of the instability strip. While even low
quality photometric data will yield a period precise to at least a few seconds. For the hotter \zzc stars,
with little or no amplitude modulation, the uncertainty in WMP
constitutes less than 1\% of the width of the instability strip in period.
Cooler \zzc stars with a significant amplitude modulation can exhibit pulsation spectra with a WMP different by
30--60\,s and in a few cases even as much as 100--200\,s. For most of the cooler \zzc stars, the uncertainty 
in WMP represents less than 5\% of the width
of the instability strip\footnote{G29-38 is the only example we find in the literature
where the published pulsation spectra exhibit a change in WMP 
by 375\,s \citep[see][]{McGrawaRobinson75,Kleinman95,Kleinmanet98}.
It is a complex pulsator with a spectroscopic
temperature of a hot \zzc star \citep{Bergeronet04}; its large amplitude probably
indicates a temperature excursion of up to 500\,K, a substantial part of the instability strip.
It seems difficult to determine a reliable
effective temperature of G29-38 by either method.}.
We therefore expect that this method of using the WMP to effectively
measure the location of a DAV 
within the instability strip is in general more accurate and reliable than spectroscopy.

Any dispersion in period due to differences
in stellar mass, core composition, H/He layer masses, etc., can 
increase the uncertainty of determining the temperature of a given DAV using this relation.
We expect this may be significantly smaller than our present typical spectroscopic
$\teff$ uncertainties of a few 100\,K.

Note that we do not claim that the relationship between the WMP
and the temperature is linear; a straight line is
merely the simplest model fit possible to the
observations shown in Figure 1, considering the large amount of scatter.
We do not necessarily require a linear relation between WMP and $\teff$ for this
method to work as a temperature indicator.
Our collaborators at the University of Texas are presently investigating
the interpretation of these data in terms of the
theoretical models (Kim \etal 2006; in preparation).                                                                                               

The plot of WMP vs. $\teff$ is similar to comparing two independent
temperature scales, each with its own independent source of uncertainties.
The uncertainty in spectroscopic $\teff$ depends on the quality of the spectrum, the accuracy
and completeness of the model atmosphere, and the details of the algorithm used to fit the
observed spectrum with the template spectra. Large amplitude variables have a corresponding
higher uncertainty in temperature \citep{McGraw79}.
The uncertainty in WMP comes from
the quality of the photometric observations, and the amplitude modulation of the star.

        As long as we restrict our relative $\teff$ parameter
in units of seconds in the WMP temperature scale, the uncertainty in
our measurements is not related to spectroscopic temperature values at all.
It is only when we attempt to translate WMP into a temperature in degree Kelvin, that
we have to use the relation between WMP and spectroscopic temperature. Even in this case,
we are still better off than the typical 200\,K uncertainty in spectroscopic temperature
because the slopes of the best fit lines in Figure 1 do not depend on a few stars, but
on 35--40 stars.
Using WMP directly as a temperature scale is non-intuitive at the present time.
But this maybe worth thinking about as an alternative scale in the long run, once we
improve our understanding of the relation between WMP and $\teff$, both observationally and theoretically.

\subsubsection{Re-defining the \zzc classification}
We presently classify the \zzc stars into two groups: the hot DAVs and the cool DAVs.
However a clear dividing point (in temperature) between these two classes does not exist
to date in published literature. Our present classification of these stars is thus
vague in this context, although it has certainly proved to be a useful guide.
The primary reason such a temperature based classification cannot be well defined is because
the location and width of the instability strip are model-dependent features.
Model atmospheres of DAV stars treat convection with some parameterization, the choice of
which can shift the instability strip in temperature by a few thousand K(\citealt{Bergeronet95,KoesteraAllard00}).                                                                                                                  
                                                                                                                  
The pulsation characteristics of the \zzc stars helped us divide them into simple
and complex pulsators in the late seventies. A decade later, we recognized that the simple pulsators
that exhibit a few modes with short periods (200--300\,s), small amplitudes (few mma), sinusoidal or saw-toothed pulse shapes,
and continued to show the same pulsation spectra over a few decades, were hot \zzc stars. We also realized that the
complex pulsators that exhibit several long periods (600--1200\,s) with large amplitudes (40--110\,mma),
non-sinusoidal pulse shapes with fast rise times and slow decadence, and amplitude modulation over
timescales of a few days to a few years, were cool \zzc stars. Although we now use spectroscopic
temperature as a means to classify these stars, the classification scheme came about only because we
initially used the pulsation characteristics to separate these stars into two groups.

We suggest a new \zzc classification scheme based on the WMP
of these variables. We intend to retain the fundamental aspect of the previous scheme
in using a temperature based classification. We have shown that WMP is also a temperature scale (Figure 1).
We expect that WMP serves as a more accurate and reliable $\teff$ scale than spectroscopic temperature
because measuring the pulsation periods of a \zzc star constitutes a relatively
simple, less uncertain, and model-independent exercise
compared to measuring its spectroscopic temperature.
                                                                                                                                                                                                                                     
We redefine the class of hot DAVs (hDAVs) as \zzc stars with WMP\,$<$\,350\,s.
We redefine the class of cool DAVs (cDAVs) as \zzc stars with WMP\,$>$\,650\,s.
We suggest introducing a new class of DAVs, to be called the intermediate DAVs (iDAVs), as
\zzc stars with 350\,$\leq$\,WMP\,$\leq$\,650. This class merely forms the
evolutionary subclass adjoining the hot and cool \zzc stars and typically
encompasses those \zzc pulsators that show a large range of pulsation periods, e.g. HS0507+0435B.

For borderline cases such as G29-38, when one season of observations place the DAV in one class,
and a second season places it in another class, then we suggest choosing the coolest class
of the two possibilities. We find several cases where a cDAV or an iDAV exhibit modes typical of the hotter
pulsators, but we have not seen an instance of a hot DAV exhibiting a mode typical of the cool \zzc stars.
We show our suggested classification for most of the known non-interacting \zzc stars in Tables 1 and 2.

Note that the boundaries of 350\,s and 650\,s are arbitrary in some sense; we merely wished to divide the
instability strip into three parts and used a period histogram to fine-tune our choice.                                                                                                                    
For simplicity and better readability, Tables 1 and 2 in the appendix do not show all of the pulsation spectra we accrued
for each star. We show multiple seasons of observations only for those DAVs that exhibit different frequencies and amplitudes
at different times. This also helps us understand the intrinsic changes in weighted mean period and mean
pulsation amplitude for such variables. To use such stars in the ensemble,
we computed the weighted mean period and mean pulsation amplitude
for each season of observations individually, and then included the average value in our analysis.
We refer the reader to \citet{Silvottiet05a}, \citet{Castanheiraet05}, \citet{Kepleret05}, \citet{Gianninaset05},                                                                                                                  
and \citet{Silvottiet05b} for the pulsation spectra of the new DAVs.
                                                                                                                   
\subsection{Observed pulsation amplitudes}
The physical quantity of interest in this subsection is the intrinsic
pulsation amplitude of modes excited in the \zzc star. But 
the measured amplitudes of observed modes
are most likely lower than the intrinsic amplitudes due to geometric
cancellation. This effect has three independent causes: disk averaging,
inclination angle, and limb darkening.
We are unable to resolve the disk of the star from Earth. Hence the observed
amplitude of each pulsation mode is lower due to a disk-averaging effect. This
explains why the probability of detecting $\ell$=1 modes is higher 
than the detection probability for $\ell$=2 and $\ell$=3 modes.
The inclination angle dictates the distribution of the bright and dark
zones in our view for a given mode. Eigenmodes with
different $m$ values exhibit different cancellation patterns (see \citealt{Dziembowski77,Pesnell85}).

Limb darkening effectively reduces the area of the stellar disk in our view,
and this reduction in area depends on wavelength.
At UV wavelengths, the increased limb darkening
decreases the contribution of zones near the limb.
As a result, modes of higher
$\ell $ are canceled less effectively in the UV compared to the
low $\ell $ modes \citep{Robinsonet82}.
Of these three independent causes of geometric cancellation, limb darkening is
the only one that works in our favor, and provides us with a mode
identification technique \citep{Robinsonet95}.

The intrinsic pulsation amplitude depends on the mass of the star.
Nonradial g-modes have a non-negligible radial component, the
amplitude of which scales with stellar mass and plays a role
in dictating the amplitude of the nonradial component.
The massive pulsators BPM\,37093 ($\log~g$=8.81),
WD0923+0120 ($\log~g$=8.74), WD1711+6541 ($\log~g$=8.64), and WD2159+1322 ($\log~g$=8.61)
exhibit low amplitudes as a result of their high gravity, and hence
we exclude them from section 2.3.

Figure 2 shows the square root of total power $\sqrt{(\sum_i A_i^2)}$ for 
38 stars from the SDSS sample (top) and 38 stars 
from the BG04 \zzc sample (bottom), plotted as a function of their spectroscopic temperature.
All of the pulsation amplitudes we report in this paper come from optical whole disk observations.
We expect these are mostly low $\ell$ modes. For these reasons, we expect them to be lower
than the corresponding intrinsic amplitudes due to geometric cancellation.
The points that form the upper
envelopes in both panels of Figure 2
are then better indicators of the intrinsic amplitude at that temperature.
Note that in the few cases of under-resolved data, beating of closely 
spaced periodicities can lead us to determine a relatively smaller
or larger amplitude than the observable amplitude.

We expect that \zzc stars with a high pulsation amplitude have a corresponding
higher uncertainty in temperature. For large amplitude variables, the surface temperature changes
substantially during a pulsation cycle, by as much as a thousand degrees \citep{McGraw79}. 
Depending on which phase of the pulsation cycle (typical periods of
600--1200\,s for cool DAVs) we acquire the spectra and for how long, our 
measure of their effective temperature can be incorrect by
a few hundred degrees. We can attempt to estimate this uncertainty by obtaining
time-series photometry on the star simultaneous with the spectroscopic data.

At any given temperature, the apparently low amplitude variables
could be suffering from extensive geometric cancellation. At the same time, the high amplitude
variables with the expectedly least geometric cancellation have a proportionally 
high uncertainty in temperature due to the 
large intrinsic temperature fluctuation during a
pulsation cycle. This makes the task of interpreting Figure 2 difficult.
There is fortunately a silver lining to this bleak cloud: \zzc stars are multi-mode
pulsators. If a \zzc with 3 independent modes still exhibits a small amplitude, then it is
unlikely that we are dealing with an unfavorable inclination angle in all three cases\footnote{This may not
hold true if all the independent modes have the same values of $\ell$ and $m$. However we do not
fully understand the mode selection mechanism for different $m$ values. Also, they exhibit different cancellation 
patterns \citep{Dziembowski77,Pesnell85}.}.
\clearpage
\begin{figure}[!ht]
\figurenum{2}
\epsscale{1.0}
\plotone{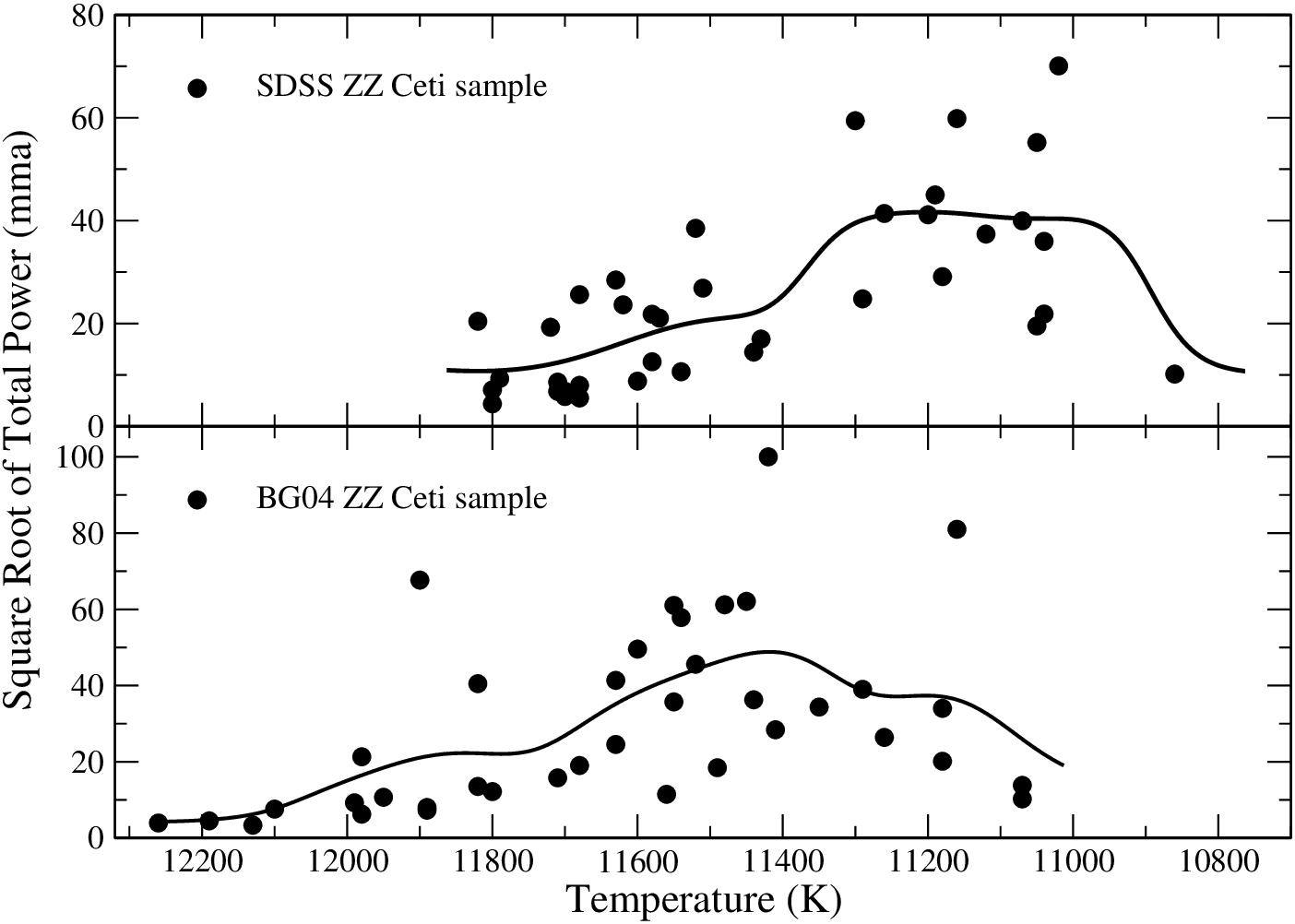}
\caption{We show the square root of total power for 38 SDSS DAVs (top) and 38 DAVs from the BG04 sample (bottom)
as a function of their spectroscopic temperature. The solid line shows a smooth curve obtained from Gaussian binning
that demonstrates an initial increase in pulsation amplitude followed by a suggestive decline
prior to the red edge.}
\end{figure}
\clearpage

Instead of traditional histograms, we adopt a Gaussian binning technique where each bin serves as a
Gaussian function instead of a box (top-hat) function. This is a better noise averaging technique
because all points contribute to the
value of each bin. Although the contributions of distant points to the value of an individual bin are small,
this method may not be appropriate to study sharp local trends. 
We are interested in slow trends across the width of the instability strip and do not hesitate to
use this technique. We choose Gaussian bins with $\sigma = 75\,K$ that are an infinitesimal 1\,K 
apart from each other. 
We show the result as a
solid line in both panels of Figure 2.
Both samples show an initial increase in pulsation
amplitude, and are also suggestive of a decline
prior to the red edge.

We established in section 2.2 how the weighted mean pulsation period correlates directly with the 
effective temperature of the DAVs for both samples. We now merge both the SDSS and the BG04
samples, also including the five DAVs with unreliable spectroscopic fits (see section 1.1) 
to plot the pulsation amplitude as a function of the WMP.
We no longer have to worry about the internal consistency
in their spectroscopic temperatures, and we can use their WMP as a $\teff $ scale.
We show these 81 DAVs as filled circles in Figure 3. We also include the 19 new DAVs
from \citet{Silvottiet05a,Silvottiet05b}, \citet{Castanheiraet05}, and \citet{Kepleret05} as filled squares.
We have excluded the massive DAV WD1337+0104 ($\log~g$=8.55) from \citet{Kepleret05}
and the low mass DAVs HE0031-5525 ($\log~g$=7.65) \& WD2135-0743 ($\log~g$=7.67) from \citet{Castanheiraet05}
in Figure 3.

Figure 3 shows a plot with better statistics
than Figure 2 due to the larger sample size of 100 DAVs.
We used Gaussian bins of width 75\,s, separated
from adjacent bins by an infinitesimal amount of 1\,s, to produce both the curves shown in
Figure 3. The solid line is the histogram determined from the 81 average mass DAVs of the BG04 
and SDSS samples, while the dotted line shows the effect of including the new 19 average mass
DAVs. The minor difference between the two lines assures us that our conclusions are robust.
We clearly see an initial increase in pulsation amplitude
near the blue edge, followed by a gentle rise and then a
decline prior to the red edge.
 \clearpage
\begin{figure}[!ht]
\figurenum{3}
\epsscale{1.0}
\plotone{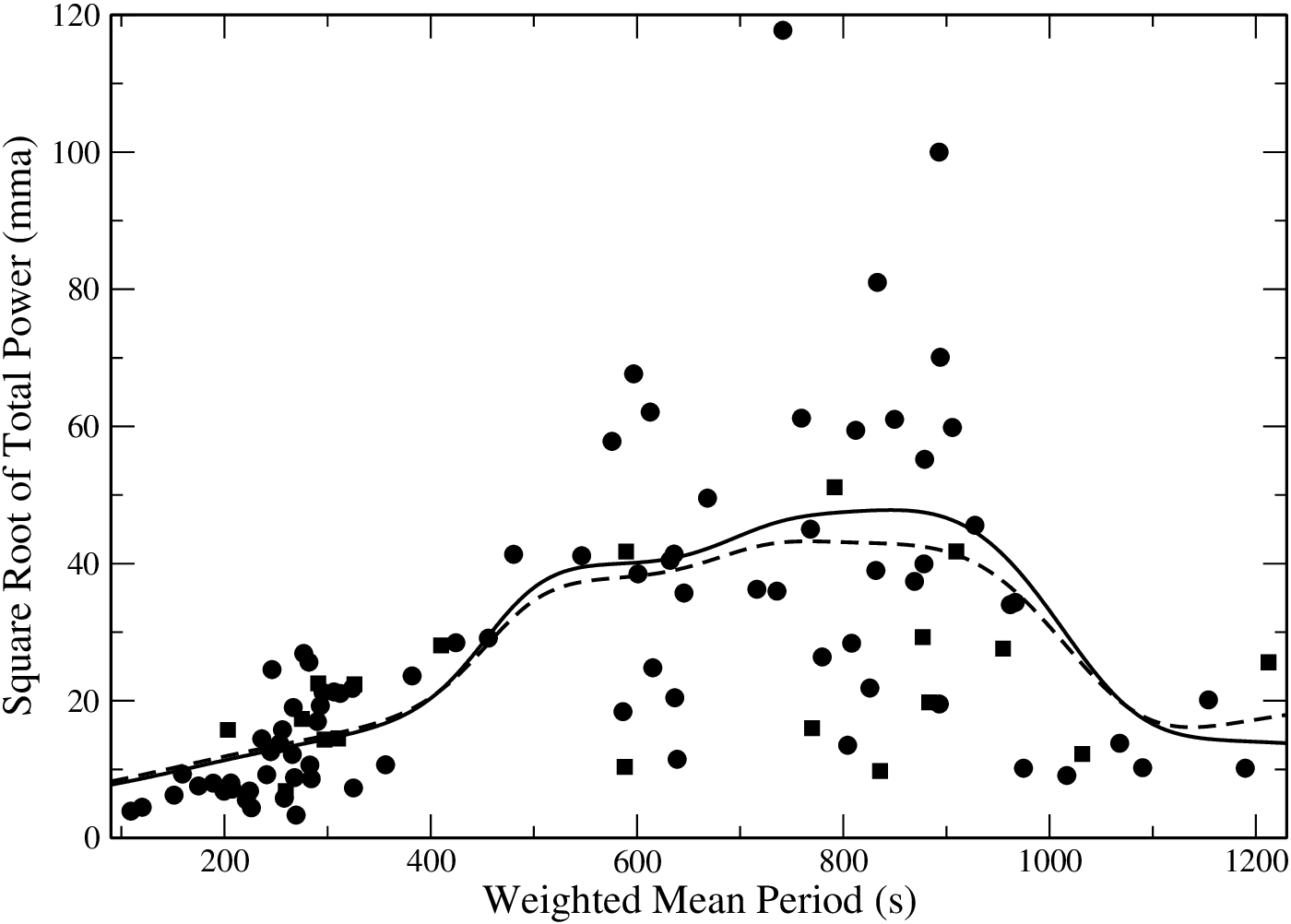}
\caption{We show the square root of total power for 81 DAVs from the SDSS and BG04 samples (filled circles)
and 19 newly discovered DAVs (filled squares) as a
function of the weighted mean period, which is correlated to the temperature (see Figure 1).
The solid line shows a histogram obtained using Gaussian bins of $\sigma $=75\,s for the 81 DAVs, and 
the dotted line shows the effect of including the 19 new DAVs.
We see an initial increase in pulsation amplitude near the blue edge, and a subsequent gentle rise followed by
a decline in amplitude prior to the red edge of the \zzc instability strip.}
\end{figure}
\clearpage
	We find several reasons why Figure 3 looks much more convincing of a decline in 
amplitude before the red edge compared to Figure 2. Changing from a x-axis based on spectroscopic temperature
to weighted mean period allows us to include four red edge pulsators with low amplitudes, namely HS0952+1816,
WD1443+0134, WD0249-0100, and WD2307-0847. Secondly, we get better statistics by combining the 
two samples that are individually
suggestive of a decline in amplitude near the red edge. Lastly, we also find that the change in x-axis
between Figures 2 and 3 moves pulsators left and right, apparently 
leaving behind cleaner evidence of
a decline prior to the red edge. For example, we find that the low amplitude cool DAV GD\,154 moves 
from 11180\,K in Figure 2 to 1154\,s in Figure 3, which corresponds
to $\teff \sim 10950$\,K using the slope in the bottom panel of Figure 1 as a conversion factor.

We have eight pulsators with weighted mean periods of order or greater than
1000\,s with amplitudes near or less than 25\,mma. There are no high amplitude pulsators in this range,
and the scatter in amplitude near the red edge is much smaller than the center of the \zzc instability strip. 
It is unlikely that unfavorable inclination angles can explain the low amplitudes of all the independent modes
of all eight pulsators.

\citet{Kanaanet02} established the observed red edge at 11000\,K, 
but did not see any decline in pulsation amplitude near the red edge within the instability strip,
although they were searching for it. 
It is for the first time that we now have clear evidence of a decline in pulsation amplitude
within the instability strip, just prior to the \zzc red edge. In other words, the star loses pulsation
energy before pulsations shut down at the red edge. This work will have implications and new constraints for
our pulsation models. 

\section{Conclusions}
We find that the present large sample of non-interacting
DAVs conforms to the well established trend of increasing pulsation periods
with decreasing temperature across the instability strip. 
Investigations of the number of observed independent modes show
that hot DAVs have one more mode on average compared to the cool DAVs.
We also confirm the increase in pulsation
amplitude near the blue edge and find strong evidence of a decline in
amplitude prior to the red edge. 
\citet{Kanaanet02} established the observed \zzc red edge at 11000\,K,
but did not see any decline in pulsation amplitude near the red edge within the instability strip.
We present the first
observational evidence that shows the red edge is not an abrupt feature in \zzc evolution, and that
the star loses pulsation energy before pulsations shut down at the red edge. This work poses
new constraints on our pulsation models.

\acknowledgements
Support for this work was provided by NASA through the Hubble Fellowship grant 
HST-HF-01175.01-A awarded by the Space Telescope Science Institute, which is operated 
by the Association of Universities for Research in Astronomy, Inc., for NASA, under contract NAS 5-26555. We thank the referee for helping us make this manuscript
a better paper.

\appendix

\clearpage
\begin{deluxetable}{llllllllllll}
\tabletypesize{\tiny}
\rotate
\tablecolumns{12}
\tablewidth{0pc}
\tablecaption{Pulsation spectra of the 41\,+\,5 ZZ Ceti stars mainly from the SDSS \zzc sample}
\tablehead{
\colhead{Object}& \colhead{Class} & \colhead{$\teff$ (K)} & \colhead{\hspace{-0.1in}$\log~g$} &\multicolumn{7}{c}{Linearly independent modes: Period (s) and Amplitude (mma)} }
\startdata
WD0018+0031\tablenotemark{\alpha}& hDAV  &   11700 & 7.93 & 257.9 5.8  & & & & & &\\
WD0048+1521\tablenotemark{\alpha} & iDAV &   11290 & 8.23 & 615.3 24.8 & & & & & &\\
WD0102$-$0032  & cDAV &  11050 & 8.24 & 1043.4 15.3 & 926.3 34.7 & 830.3 35.1 & 752.2 19.4 & & &\\
WD0111+0018  & hDAV &  11510 & 8.26 & 292.3 21.9 & 255.3 15.6 & & & & &\\
WD0214$-$0823  & hDAV &  11570 & 7.92 & 348.1 8.4 & 347.1 8.2 & 297.5 16.0 & 263.5 7.1 & & & \\
WD0318+0030  & cDAV &  11040 & 8.07 & 844.9 15.3 & 826.4 27.3 & 695.0 8.9 & 587.1 10.6 & 536.1 11.1 & & \\
WD0332$-$0049  & cDAV &  11040 & 8.25 & 1143.7 7.4 & 938.4 6.7 & 765.0 15.2 & 402.0 4.1 & & & \\
WD0332$-$0049  & cDAV &   11040 & 8.25 & 770 23 & 910 10& & & & & \\
WD0756+2020\tablenotemark{\alpha}  & hDAV &  11710 & 8.01 & 199.5 6.8& & & & & & \\
WD0815+4437  & iDAV &  11620 & 7.93 & 787.5 7.2 & 311.7 18.9 & 309.0 10.2 & 258.3 6.8& & & \\
WD0818+3131\tablenotemark{\alpha}  & hDAV &  11800 & 8.07 & 253.3 2.9 & 202.3 3.3& & & & & \\
WD0825+4119  & iDAV &  11820 & 8.49 & 653.4 17.1 & 611.0 11.2& & & & \\
WD0842+3707  & hDAV &  11720 & 7.73 & 321.1 4.4 & 309.3 18.0 & 212.3 5.2 & & & & & \\
WD0847+4510  & hDAV &  11680 & 8.00 & 200.5 7.0 & 123.4 3.0& & & & & \\
WD0906$-$0024  & iDAV &  11520 & 8.00 & 769.4 26.1 & 618.8 9.3 & 574.5 23.7 & 457.9 9.8 & 266.6 7.6& & \\
WD0913+4036\tablenotemark{\alpha}  & hDAV &  11680 & 7.87 & 320.5 14.7 & 288.7 12.4 & 260.3 16.5 & 203.9 3.8& & & \\
WD0923+0120  & cDAV &  11150 & 8.74 & 655.7 4.4& & & & & & & \\
WD0923+0120  & cDAV &  11150 & 8.74 & 668.9 3.5& & & & & & & \\
WD0939+5609  & hDAV &  11790 & 8.22 & 249.9 7.2 & 48.5 5.9& & & & & & \\
WD0942+5733  & iDAV &  11260 & 8.27 & 909.4 7.7 & 694.7 37.7 & 550.5 12.3 & 273.0 9.0 & & & & \\
WD0949$-$0000  & iDAV &  11180 & 8.22 & 711.6 6.0 & 634.2 5.1 & 516.6 16.2 & 365.2 17.7 & 364.1 7.3 & 363.2 12.5 & & \\
WD0958+0130  & hDAV &  11680 & 7.99 & 264.4 4.7 & 203.7 2.5 & 121.2 1.6& & & & & \\
WD1002+5818\tablenotemark{\alpha}  & hDAV &  11710 & 7.92 & 304.6 5.3 & 268.2 6.8& & & & & \\
WD1007+5245\tablenotemark{\alpha}  & hDAV &  11430 & 8.08 & 323.1 10.4 & 290.1 7.7 & 258.8 11.0& & & & \\
WD1015+0306  & hDAV &  11580 & 8.14 & 270.0 8.4 & 255.7 7.3 & 194.7 5.8& & & & & \\
WD1015+5954  & iDAV &  11630 & 8.02 & 769.9 5.7 & 453.8 15.5 & 401.7 21.4 & 292.4 8.6& & & & \\
WD1015+5954  & iDAV &  11630 & 8.02 & 768.4 7.5 & 455.3 17.7 & 401.4 19.8 & 294.0 9.1& & & & \\
WD1015+5954  & iDAV &  11630 & 8.02 & 456.1 19.6 & 399.7 19.2 & 145.5 3.1 & 139.2 4.6 & & & & \\
WD1054+5307\tablenotemark{\alpha}  & cDAV &  11120 & 8.01 & 869.1 37.4 & & & & & & & \\
WD1056$-$0006  &  cDAV & 11020 & 7.86 & 1024.9 31.6 & 925.4 60.3 & 670.6 12.0 & 603.0 11.5 & & & & \\
WD1122+0358  & cDAV &  11070 & 8.06 & 996.1 17.3 & 859.1 34.6 & 740.1 10.0& & & & & \\
WD1125+0345  &  hDAV & 11600 & 7.99 & 335.1 2.8 & 265.8 3.3 & 265.5 7.1 & 208.6 2.8& & & & \\
WD1157+0553  & cDAV &  11050 & 8.15 & 1056.2 5.8 & 918.9 15.9 & 826.2 8.1 & 748.5 5.6& & & \\
WD1345$-$0055  & hDAV &  11800 & 8.04 & 254.4 2.4 & 195.5 3.9 & 195.2 5.5& & & & & \\
WD1354+0108  & hDAV &  11700 & 8.00 & 322.9 1.9 & 291.6 2.2 & 198.3 6.0 & 173.3 1.1 & 127.8 1.5& & \\
WD1355+5454\tablenotemark{\alpha}  & hDAV &  11580 & 7.95 & 324.0 21.8& & & & & & \\
WD1417+0058  & cDAV &  11300 & 8.04 & 980.0 11.3 & 894.6 42.8 & 812.5 32.1 & 749.4 17.9 & 522.0 14.9& & \\
WD1502$-$0001  & iDAV &  11200 & 8.00 & 687.5 12.0 & 629.5 32.6 & 581.9 11.1 & 418.2 14.9 & & & & \\
WD1617+4324  & cDAV &  11190 & 8.03 & 889.7 36.6 & 661.7 21.2 & 626.3 15.4 & & & & & \\
WD1700+3549  & cDAV &  11160 & 8.04 & 1164.4 11.4 & 955.3 20.3 & 893.4 54.3 & 552.6 9.3 & & & \\
WD1711+6541  & cDAV &  11310 & 8.64 & 1248.2 3.2 & 690.2 3.3 & 606.3 5.2 & 234.0 1.2& & & & \\
WD1711+6541  & cDAV &  11310 & 8.64 & 1186.6 3.3 & 934.8 2.9 & 612.6 5.7 & 561.5 3.0& 214.3 1.7& & & \\
WD1724+5835  & hDAV &  11540 & 7.89 & 337.9 5.9 & 279.5 8.3 & 189.2 3.2& & & & & \\
WD1732+5905  & cDAV &  10860 & 7.99 & 1248.4 4.5 & 1122.4 8.0 & & & & & & \\
WD1732+5905  & cDAV &  10860 & 7.99 & 1336 7.8 & 1090 8.0& & & & & & \\
WD2159+1322\tablenotemark{\alpha}  & cDAV &  11710 & 8.61 & 801.0 15.1 & 683.7 11.7& & & & & & \\
WD2214$-$0025\tablenotemark{\alpha}  & hDAV &  11440 & 8.33 & 255.2 13.1 & 195.2 6.1& & & & & & \\
\hline
\hline
WD1443+0134  & cDAV &  10830\tablenotemark{\beta} & \nodata & 1085.0 5.2 & 969.0 7.5& & & & & & \\
WD1524$-$0030  & cDAV &  11160\tablenotemark{\beta} & \nodata & 873.3 110.8 & 717.5 28.3 & 498.6 21.6 & 255.2 17.9& & & \\
WD2350$-$0054  & hDAV &  11690\tablenotemark{\beta} & \nodata & 391.1 3.1 & 304.1 16.3 & 272.8 16.2 & & & & & \\
WD2350$-$0054  & hDAV &  11710\tablenotemark{\beta} & \nodata & 304.5 13.8 & 272.7 14.8 & 212.6 3.0& & & & & \\
WD2350$-$0054  & hDAV &  11680\tablenotemark{\beta} & \nodata & 391.1 7.5 & 304.3 17.0 & 273.3 6.3 & 206.7 3.2& & & & \\
HS0951+1312  & hDAV &  11740\tablenotemark{\beta} & \nodata & 311.7 2.7 & 282.2 9.0 & 258.0 3.6 & 208.0 9.4& & & & \\
HS0952+1816  & cDAV &  10800\tablenotemark{\beta} & \nodata & 1160.9 7.9 & 945.9 10.4& & & & & & \\
HS0952+1816  & cDAV &  10960\tablenotemark{\beta} & \nodata & 1150 4.8 & 883 3.6 & 790 2.9 & 674.7 3.0& & & & \\
\enddata
\tablenotetext{\alpha}{We obtained the pulsation spectrum from \citet{Mullallyet05}.}
\tablenotetext{\beta }{We derived these temperatures using the best fit to the weighted mean period\,--\,spectroscopic 
temperature plot (Figure 1, top panel) using the first 41 SDSS DAVs listed in this table.} 
\end{deluxetable}
\clearpage
\setlength{\footskip}{-2in}
\thispagestyle{empty}
\begin{deluxetable}{llllllllllllllll}
\tabletypesize{\tiny}
\rotate
\tablecolumns{16}
\tablewidth{0pc}
\tablecaption{Pulsation spectra of the 36 ZZ Ceti stars in the BG04 \zzc sample}
\tablehead{
\colhead{\hspace{-0.3in}Object}& \colhead{\hspace{-0.15in}Class} & \colhead{\hspace{-0.15in}$\teff$\,(K)} & \colhead{\hspace{-0.1in}$\log~g$} &\multicolumn{11}{c}{Linearly independent modes: Period (s) and Amplitude (mma)} &\colhead{\hspace{-0.1in}Reference}}
\startdata
\hspace{-0.3in}BPM30551 & \hspace{-0.1in}cDAV       & \hspace{-0.1in}11260 & \hspace{-0.1in}8.23 & 920.5 18.5 & 741.4 21.7 & 655.4 17.4 & 442.8 6.5 & & & & & & & & \hspace{-0.1in}\citealt{HesserLaskeraNeupert76}\\
\hspace{-0.3in}BPM30551 & \hspace{-0.1in}cDAV       & \hspace{-0.1in}11260 & \hspace{-0.1in}8.23 & 993.0 6.5 & 936.2 13.0 & 885.6 15.2 & 819.2 19.5 & 738.0 6.5 & 609.6 6.5 & & & & & & \hspace{-0.1in}\citealt{HesserLaskeraNeupert76}\\
\hspace{-0.3in}BPM30551  & \hspace{-0.1in}cDAV      & \hspace{-0.1in}11260 & \hspace{-0.1in}8.23 & 963.8 13.0 & 844.5 19.5 & 799.2 10.9 & 682.7 13.0 & 606.8 13.0 & & & & & & & \hspace{-0.1in}\citealt{HesserLaskeraNeupert76}\\
\hspace{-0.3in}BPM30551   & \hspace{-0.1in}cDAV     & \hspace{-0.1in}11260 & \hspace{-0.1in}8.23 & 920.4 8.7 & 751.6 13.0 & 682.7 10.9 & 606.8 15.2 & 546.1 5.4 & 496.5 4.3 & & & & & & \hspace{-0.1in}\citealt{HesserLaskeraNeupert76}\\
\hspace{-0.3in}BPM30551    & \hspace{-0.1in}cDAV    & \hspace{-0.1in}11260 & \hspace{-0.1in}8.23 & 910.2 8.7 & 862.3 6.5 & 731.4 6.5 & 682.7 13.0 & 607.9 17.4 & & & & & & & \hspace{-0.1in}\citealt{HesserLaskeraNeupert76}\\
\hspace{-0.3in}BPM30551  & \hspace{-0.1in}cDAV      & \hspace{-0.1in}11260 & \hspace{-0.1in}8.23 & 1129.9 10.9 & 1057.0 8.7 & 744.7 18.5 & 606.8 16.3 & & & & & & & & \hspace{-0.1in}\citealt{HesserLaskeraNeupert76}\\
\hspace{-0.3in}BPM30551  & \hspace{-0.1in}cDAV      & \hspace{-0.1in}11260 & \hspace{-0.1in}8.23 & 1137.8 8.7 & 958.1 7.6 & 862.3 16.3 & 744.7 7.6 & 606.8 14.1 & & & & & & & \hspace{-0.1in}\citealt{HesserLaskeraNeupert76}\\
\hspace{-0.3in}BPM30551  & \hspace{-0.1in}cDAV      & \hspace{-0.1in}11260 & \hspace{-0.1in}8.23 & 1092.3 5.4 & 993.0 8.7 & 936.2 9.8 & 744.7 7.6 & 712.3 6.5 & 606.8 11.9 & & & & & & \hspace{-0.1in}\citealt{HesserLaskeraNeupert76}\\
\hspace{-0.3in}BPM30551 & \hspace{-0.1in}cDAV       & \hspace{-0.1in}11260 & \hspace{-0.1in}8.23 & 862.3 6.5 & 799.2 8.7 & 744.7 7.6 & 668.7 7.6 & 612.5 5.4 & & & & & & & \hspace{-0.1in}\citealt{HesserLaskeraNeupert76}\\
\hspace{-0.3in}R548   & \hspace{-0.1in}hDAV         & \hspace{-0.1in}11990 & \hspace{-0.1in}7.97 & 333.6 0.6 & 318.0 0.9 & 274.8 3.8 & 274.3 4.8 & 213.1 7.4 & 212.8 4.7 & 187.3 0.9 & & & & & \hspace{-0.1in}\citealt{Mukadamet03}\\
\hspace{-0.3in}MCT0145-2211  & \hspace{-0.1in}iDAV  & \hspace{-0.1in}11550 & \hspace{-0.1in}8.14 & 823.2 17 & 727.9 19 & 462.2 25 & & & & & & & & & \hspace{-0.1in}\citealt{Fontaineet03}\\
\hspace{-0.3in}BPM31594   & \hspace{-0.1in}iDAV     & \hspace{-0.1in}11540 & \hspace{-0.1in}8.11 & 617.9 48.0 & 401.6 16 & 416.1 5 & & & & & & & & & \hspace{-0.1in}\citealt{ODonoghueWarneraCropper92}\\
\hspace{-0.3in}HLTau76  & \hspace{-0.1in}iDAV       & \hspace{-0.1in}11450 & \hspace{-0.1in}7.89 & 933 25.2 & 796 9.7 & 781 9.9 & 541 40.5 & 494 30.2 & 383 21.8 & & & & & & \hspace{-0.1in}\citealt{Dolez98}\\
\hspace{-0.3in}G38-29    & \hspace{-0.1in}cDAV      & \hspace{-0.1in}11180 & \hspace{-0.1in}7.91 & 1024.0 26.1 & 938.0 26.5 & & & & & & & & & & \hspace{-0.1in}\citealt{McGrawaRobinson75}\\
\hspace{-0.3in}G38-29     & \hspace{-0.1in}cDAV     & \hspace{-0.1in}11180 & \hspace{-0.1in}7.91 & 1019.8 12.2 & 910.3 28.3 & & & & & & & & & & \hspace{-0.1in}\citealt{McGrawaRobinson75}\\
\hspace{-0.3in}G191-16    & \hspace{-0.1in}cDAV     & \hspace{-0.1in}11420 & \hspace{-0.1in}8.05 & 892.9 100 & & & & & & & & & & & \hspace{-0.1in}\citealt{Vauclairet89}\\
\hspace{-0.3in}HS0507+0435B & \hspace{-0.1in}iDAV   & \hspace{-0.1in}11630 & \hspace{-0.1in}8.17 & 743.0 13.9 & 557.7 18.7 & 446.2 11.0 & 444.8 13.6 & 355.8 22.7 & 354.9 6.8 & 286.1 3.6 & & & & & \hspace{-0.1in}\citealt{KotakKerkwijkaClemens02}\\
\hspace{-0.3in}HS0507+0435B  & \hspace{-0.1in}iDAV & \hspace{-0.1in}11630 & \hspace{-0.1in}8.17 & 743.4 8.3 & 588.7 3.9 & 583.8 1.6 & 559.6 2.7 & 557.6 17.4 & 557.2 3.1 & 556.5 7.8 & \hspace{-0.1in}555.3 18.0 & \hspace{-0.1in}446.1 15.1 & \hspace{-0.1in}445.3 3.0 & \hspace{-0.1in}444.6 12.9 & \hspace{-0.1in}\citealt{Handleret02}\\
\hspace{-0.3in} &        &       &       & 355.8 26.1 & 355.4 4.7 & 354.9 11.1 & & & & & & & & & \\
\hspace{-0.3in}GD66     & \hspace{-0.1in}hDAV       & \hspace{-0.1in}11980 & \hspace{-0.1in}8.05 & 649.4 2.2 & 441.9 1.6 & 301.7 6.7 & 271.7 8.4 & 196.5 3.6 & & & & & & & \hspace{-0.1in}\citealt{Fontaineet85}\\
\hspace{-0.3in}GD66      & \hspace{-0.1in}hDAV      & \hspace{-0.1in}11980 & \hspace{-0.1in}8.05 & 788.6 5.5 & 461.0 3.6 & 301.7 6.3 & 271.7 30 & 197.2 2.2 & & & & & & & \hspace{-0.1in}\citealt{Fontaineet85}\\
\hspace{-0.3in}GD66    & \hspace{-0.1in}hDAV        & \hspace{-0.1in}11980 & \hspace{-0.1in}8.05 & 304.5 8.8 & 271.1 14.8 & 256.5 9 & 197.8 7 & 123.1 2.3 & & & & & & & \hspace{-0.1in}\citealt{Fontaineet01}\\
\hspace{-0.3in}HE0532-5605  & \hspace{-0.1in}iDAV   & \hspace{-0.1in}11560 & \hspace{-0.1in}8.49 & 688.8 8.3 & 586.4 7.9 & & & & & & & & & & \hspace{-0.1in}\citealt{Fontaineet03}\\
\hspace{-0.3in}KUV08368+4026 & \hspace{-0.1in}iDAV  & \hspace{-0.1in}11490 & \hspace{-0.1in}8.05 & 618.0 17.4 & 494.5 6.0 & & & & & & & & & & \hspace{-0.1in}\citealt{Vauclairet97}\\
\hspace{-0.3in}GD99      & \hspace{-0.1in}cDAV      & \hspace{-0.1in}11820 & \hspace{-0.1in}8.08 & 1151.0 1.9 & 1088.0 4.3 & 1058.0 8.3 & 1007.0 6.5 & 976.0 2.1 & 924.7 1.7 & 853.2 2.4 & \hspace{-0.1in}633.1 2.0 & \hspace{-0.1in}228.9 4.5 & \hspace{-0.1in}223.6 2.9 & \hspace{-0.1in}105.2 2.0 & \hspace{-0.1in}Chynoweth\,\&\,Thompson\,2005;\,prv.\,comm.\\
\hspace{-0.3in}G117-B15A    & \hspace{-0.1in}hDAV   & \hspace{-0.1in}11630 & \hspace{-0.1in}7.97 & 304.4 8.2 & 271.0 7.3 & 215.2 23.9 & & & & & & & & & \hspace{-0.1in}\citealt{Kepleret82}\\
\hspace{-0.3in}KUV11370+4222  & \hspace{-0.1in}hDAV & \hspace{-0.1in}11890 & \hspace{-0.1in}8.06 & 462.9 3.5 & 292.2 2.7 & 257.2 5.8 & & & & & & & & & \hspace{-0.1in}\citealt{Vauclairet97}\\
\hspace{-0.3in}G255-2    & \hspace{-0.1in}cDAV      & \hspace{-0.1in}11440 & \hspace{-0.1in}8.17 & 898.5 9.4 & 855.4 11.2 & 819.7 11.3 & 775.2 15.2 & 681.2 24.9 & 607.9 13.1 & 568.5 6.6 & & & & & \hspace{-0.1in}Vauclair\,2005;\,prv.\,comm.\\
\hspace{-0.3in}G255-2    & \hspace{-0.1in}cDAV      & \hspace{-0.1in}11440 & \hspace{-0.1in}8.17 & 985.2 4.8 & 773.4 12.7 & 681.2 27.7 & 568.5 16.5 & & & & & & & & \hspace{-0.1in}Vauclair\,2005;\,prv.\,comm.\\
\hspace{-0.3in}BPM37093   & \hspace{-0.1in}iDAV     & \hspace{-0.1in}11730 & \hspace{-0.1in}8.81 & 636.7 1.7 & 633.2 1.1 & 613.5 1.1 & 600.7 0.9 & 582.0 1.0 & 565.5 1.2 & 562.6 0.9 & \hspace{-0.1in}548.4 1.1 & & & & \hspace{-0.1in}\citealt{Kanaanet05}\\
\hspace{-0.3in}BPM37093    & \hspace{-0.1in}iDAV    & \hspace{-0.1in}11730 & \hspace{-0.1in}8.81 & 660.8 0.5 & 637.2 0.7 & 633.5 1.3 & 565.9 0.5 & 549.2 0.8 & 531.1 1.2 & 511.7 0.7 & & & & & \hspace{-0.1in}\citealt{Kanaanet05}\\
\hspace{-0.3in}HE1258+0123  & \hspace{-0.1in}cDAV   & \hspace{-0.1in}11410 & \hspace{-0.1in}8.04 & 1092.1 14 & 744.6 23 & 528.5 9 & & & & & & & & & \hspace{-0.1in}\citealt{Bergeronet04}\\
\hspace{-0.3in}GD154      & \hspace{-0.1in}cDAV     & \hspace{-0.1in}11180 & \hspace{-0.1in}8.15 & 1190.5 6.3 & 1186.5 16.7 & 1183.5 4.6 & 1092.1 3.0 & 1088.6 5.0 & 1084.0 5.6 & & & & & & \hspace{-0.1in}\citealt{Pfeifferet95}\\
\hspace{-0.3in}LP133-144   & \hspace{-0.1in}hDAV    & \hspace{-0.1in}11800 & \hspace{-0.1in}7.87 & 327.3 4 & 306.9 4 & 304.5 4 & 209.2 10 & & & & & & & & \hspace{-0.1in}\citealt{Bergeronet04}\\
\hspace{-0.3in}G238-53    & \hspace{-0.1in}hDAV     & \hspace{-0.1in}11890 & \hspace{-0.1in}7.91 & 206.2 8 & & & & & & & & & & & \hspace{-0.1in}\citealt{FontaineaWesemael84}\\
\hspace{-0.3in}EC14012-1446 & \hspace{-0.1in}iDAV   & \hspace{-0.1in}11900 & \hspace{-0.1in}8.16 & 937 11 & 610 57 & 724 21 & 530 15 & 399 13 & & & & & & & \hspace{-0.1in}\citealt{Stobieet95}\\
\hspace{-0.3in}GD165      & \hspace{-0.1in}hDAV     & \hspace{-0.1in}11980 & \hspace{-0.1in}8.06 & 249.7 0.7 & 192.8 0.9 & 192.7 2.4 & 192.6 1.9 & 166.2 0.4 & 146.4 0.5 & 120.4 1.8 & \hspace{-0.1in}120.36 4.8 & \hspace{-0.1in}120.3 1.4 & \hspace{-0.1in}114.3 0.6 & \hspace{-0.1in}107.7 0.4 &\hspace{-0.1in}\citealt{Bergeronet93}\\
\hspace{-0.3in}L19-2      & \hspace{-0.1in}hDAV     & \hspace{-0.1in}12100 & \hspace{-0.1in}8.21 & 350.1 1.1 & 348.7 0.5 & 192.6 6.5 & 193.1 0.9 & 192.1 0.8 & 143.4 0.6 & 143.0 0.3 & \hspace{-0.1in}118.9 0.3 &\hspace{-0.1in}118.7 1.2 & \hspace{-0.1in}118.5 2.0 & & \hspace{-0.1in}\citealt{ODonoghueaWarner82} \\
\hspace{-0.3in}   &      &       &       & 114.2 0.3 & 113.8 2.4 & 113.3 0.6 & & & & & & & & \\
\hspace{-0.3in}PG1541+651   & \hspace{-0.1in}cDAV   & \hspace{-0.1in}11600 & \hspace{-0.1in}8.10 & 757 14 & 689 45 & 564 15 & 467 3 & & & & & & & & \hspace{-0.1in}\citealt{Vauclairet00}\\
\hspace{-0.3in}R808        & \hspace{-0.1in}cDAV    & \hspace{-0.1in}11160 & \hspace{-0.1in}8.04 & 833 81 & & & & & & & & & & & \hspace{-0.1in}\citealt{McGrawaRobinson76}\\
\hspace{-0.3in}G226-29    & \hspace{-0.1in}hDAV     & \hspace{-0.1in}12270 & \hspace{-0.1in}8.28 & 109.5 2.8 & 109.3 1.1 & 109.1 2.5 & & & & & & & & & \hspace{-0.1in}\citealt{Kepleret95}\\
\hspace{-0.3in}BPM24754   & \hspace{-0.1in}cDAV     & \hspace{-0.1in}11070 & \hspace{-0.1in}8.03 & 1176 22.6 & & & & & & & & & & & \hspace{-0.1in}\citealt{Giovanniniet98}\\
\hspace{-0.3in}BPM24754    & \hspace{-0.1in}cDAV    & \hspace{-0.1in}11070 & \hspace{-0.1in}8.03 & 1050 9.1 & & & & & & & & & & & \hspace{-0.1in}\citealt{Giovanniniet98}\\
\hspace{-0.3in}BPM24754     & \hspace{-0.1in}cDAV   & \hspace{-0.1in}11070 & \hspace{-0.1in}8.03 & 1086 13.2 & & & & & & & & & & & \hspace{-0.1in}\citealt{Giovanniniet98}\\
\hspace{-0.3in}BPM24754    & \hspace{-0.1in}cDAV    & \hspace{-0.1in}11070 & \hspace{-0.1in}8.03 & 978 7.7 & & & & & & & & & & & \hspace{-0.1in}\citealt{Giovanniniet98}\\
\hspace{-0.3in}BPM24754   & \hspace{-0.1in}cDAV     & \hspace{-0.1in}11070 & \hspace{-0.1in}8.03 & 1098 6.1 & & & & & & & & & & & \hspace{-0.1in}\citealt{Giovanniniet98}\\
\hspace{-0.3in}BPM24754    & \hspace{-0.1in}cDAV    & \hspace{-0.1in}11070 & \hspace{-0.1in}8.03 & 1122 6.7 & & & & & & & & & & & \hspace{-0.1in}\citealt{Giovanniniet98}\\
\hspace{-0.3in}G207-9      & \hspace{-0.1in}iDAV    & \hspace{-0.1in}11950 & \hspace{-0.1in}8.35 & 557.4 6.3 & 318.0 6.4 & 292.0 5.0 & 259.1 1.7 & & & & & & & & \hspace{-0.1in}\citealt{RobinsonaMcGraw76}\\
\hspace{-0.3in}G185-32    & \hspace{-0.1in}hDAV     & \hspace{-0.1in}12130 & \hspace{-0.1in}8.05 & 651.7 0.7 & 537.6 0.6 & 454.6 0.4 & 370.2 1.6 & 301.4 1.1 & 299.8 1.0 & 264.2 0.5 & \hspace{-0.1in}215.7 1.9 & \hspace{-0.1in}141.9 1.4 & \hspace{-0.1in}72.9 0.4 & & \hspace{-0.1in}\citealt{Castanheiraet04}\\
\hspace{-0.3in}G185-32   & \hspace{-0.1in}hDAV      & \hspace{-0.1in}12130 & \hspace{-0.1in}8.05 & 370.2 1.3 & 301.6 1.5 & 215.7 1.9 & 141.9 1.5 & 72.6 0.7 & & & & & & & \hspace{-0.1in}\citealt{Thompsonet04}\\
\hspace{-0.3in}GD385     & \hspace{-0.1in}hDAV      & \hspace{-0.1in}11710 & \hspace{-0.1in}8.04 & 256.3 10.9 & 256.1 11.4 & & & & & & & & & & \hspace{-0.1in}\citealt{Kepler84}\\
\hspace{-0.3in}GD244      & \hspace{-0.1in}hDAV     & \hspace{-0.1in}11680 & \hspace{-0.1in}8.08 & 307.0 14 & 294.6 5.5 & 256.3 5 & 203.3 10.5 & & & & & & & & \hspace{-0.1in}\citealt{Fontaineet01}\\
\hspace{-0.3in}PG2303+243  & \hspace{-0.1in}cDAV    & \hspace{-0.1in}11480 & \hspace{-0.1in}8.09 & 900.5 16 & 794.5 56 & 675.4 8 & 623.4 15 & 570.7 8 & & & & & & & \hspace{-0.1in}\citealt{VauclairDolezaChevreton87}\\
\hspace{-0.3in}G29-38      & \hspace{-0.1in}cDAV    & \hspace{-0.1in}11820 & \hspace{-0.1in}8.14 & 1015.5 14.5 & 930.9 25.7 & 824.7 20.2 & 677.0 17.6 & 612.9 20.0 & & & & & & & \hspace{-0.1in}\citealt{McGrawaRobinson75}\\
\hspace{-0.3in}G29-38      & \hspace{-0.1in}cDAV    & \hspace{-0.1in}11820 & \hspace{-0.1in}8.14 & 934.5 20.5 & 813.8 23.5 & 671.5 23.0 & 623.8 11.8 & & & & & & & & \hspace{-0.1in}\citealt{McGrawaRobinson75}\\
\hspace{-0.3in}G29-38      & \hspace{-0.1in}cDAV    & \hspace{-0.1in}11820 & \hspace{-0.1in}8.14 & 859.6 24.6 & 648.7 7.8 & 614.3 31.3 & 498.3 5.8 & 401.3 4.4 & 400.4 7.0 & 399.6 8.8 & \hspace{-0.1in}283.9 3.5& & & & \hspace{-0.1in}\citealt{Kleinman95}\\
\hspace{-0.3in}G29-38       & \hspace{-0.1in}cDAV   & \hspace{-0.1in}11820 & \hspace{-0.1in}8.14 & 915.4 5.9 & 615.1 58.0 & 500.4 8.0 & 474.9 4.8 & 401.3 5.6 & 400.4 5.8 & 399.6 10.0 & \hspace{-0.1in}354.9 2.9 & \hspace{-0.1in}333.9 2.7 & \hspace{-0.1in}283.9 3.5 & \hspace{-0.1in}110.1 0.9 &\hspace{-0.1in}\citealt{Kleinman95}\\
\hspace{-0.3in}G29-38       & \hspace{-0.1in}cDAV   & \hspace{-0.1in}11820 & \hspace{-0.1in}8.14 & 770.7 1.5 & 503.5 8.6 & 495.0 11.8 & 401.2 9.7 & 400.5 1.3 & 399.7 4.9 & 283.9 6.4 & \hspace{-0.1in}177.1 0.8 & & & & \hspace{-0.1in}\citealt{Kleinman95}\\
\hspace{-0.3in}G29-38       & \hspace{-0.1in}cDAV   & \hspace{-0.1in}11820 & \hspace{-0.1in}8.14 & 809.4 30.1 & 610.3 10.6 & 401.2 11.2 & 399.7 4.5 & 283.9 6.9 & & & & & & & \hspace{-0.1in}\citealt{Kleinman95}\\
\hspace{-0.3in}G29-38       & \hspace{-0.1in}cDAV   & \hspace{-0.1in}11820 & \hspace{-0.1in}8.14 & 894.0 14.0 & 770.8 8.7 & 678.4 9.7 & 612.4 31.6 & 610.7 8.2 & 608.9 8.5 & 551.9 4.4 & \hspace{-0.1in}498.3 6.1 & \hspace{-0.1in}401.2 6.0 & \hspace{-0.1in}399.7 5.7 & \hspace{-0.1in}237.0 1.8 &\hspace{-0.1in}\citealt{Kleinman95}\\
\hspace{-0.3in}     &    &       &      & 236.4 1.8 & & & & & & & & \\
\hspace{-0.3in}G30-20     & \hspace{-0.1in}cDAV     & \hspace{-0.1in}11070 & \hspace{-0.1in}7.95 & 1068 13.8 & & & & & & & & & & & \hspace{-0.1in}\citealt{Mukadamet02}\\
\hspace{-0.3in}EC23487-2424  & \hspace{-0.1in}cDAV  & \hspace{-0.1in}11520 & \hspace{-0.1in}8.10 & 993.0 37.7 & 989.3 11.0 & 868.2 12.8 & 804.5 19.3 & & & & & & & & \hspace{-0.1in}\citealt{Stobieet93}\\
\enddata
\end{deluxetable}
\clearpage


\begin{thebibliography}{}

\bibitem[Bergeron et al.(1993)]{Bergeronet93} Bergeron, P., et al.\ 
1993, \aj, 106, 1987 

\bibitem[Bergeron et al.(1995)]{Bergeronet95} Bergeron, P., Wesemael, F., Lamontagne, R., Fontaine, G., Saffer, R.~A., \& Allard, N.~F.\ 1995, \apj, 449, 258
                                                                                                   
\bibitem[Bergeron \& Lamontagne(2003)]{BergeronaLamontagne03} Bergeron, P., 
\& Lamontagne, R.\ 2003, NATO ASIB Proc.~105: White Dwarfs, 219 

\bibitem[Bergeron et al.(2004)]{Bergeronet04} Bergeron, P.,
Fontaine, G., Bill{\` e}res, M., Boudreault, S., \& Green, E.~M.\ 2004,
\apj, 600, 404

\bibitem[Brassard et al.(1995)Brassard, Fontaine, \& Wesemael]{Brassardet95} Brassard, P., 
Fontaine, G., \& Wesemael, F.\ 1995, \apjs, 96, 545 
 
\bibitem[Brickhill(1992)]{Brickhill92} Brickhill, A.~J.\ 1992, 
\mnras, 259, 529 
 
\bibitem[Castanheira et al.(2004)]{Castanheiraet04} Castanheira, B.~G., 
et al.\ 2004, \aap, 413, 623 

\bibitem[Castanheira et al.(2005)]{Castanheiraet05} Castanheira, B.~G.,
et al.\ 2005, \aap, submitted

\bibitem[Clemens (1993)]{Clemens93} Clemens, J.\ C.\ 1993,  Ph.D. Thesis, University of Texas at Austin

\bibitem[Cox(1980)]{Cox80} Cox, J.~P.\ 1980, Research
supported by the National Science Foundation Princeton, NJ, Princeton
University Press, 1980.~393 p.

\bibitem[Dolez(1998)]{Dolez98} Dolez, N.\ 1998, Baltic 
Astronomy, 7, 153 

\bibitem[Dziembowski(1977)]{Dziembowski77} Dziembowski, W.\ 1977, 
Acta Astronomica, 27, 203 

\bibitem[Fleming et al.(1986)Fleming, Liebert, \& Green]{Fleminget86} Fleming,
T.~A., Liebert, J., \& Green, R.~F.\ 1986, \apj, 308, 176

\bibitem[Fontaine et al.(1982)]{Fontaineet82} Fontaine, G., Lacombe,
P., McGraw, J.~T., Dearborn, D.~S.~P., \& Gustafson, J.\ 1982, \apj, 258,
651

\bibitem[Fontaine \& Wesemael(1984)]{FontaineaWesemael84} Fontaine, G., \& 
Wesemael, F.\ 1984, \aj, 89, 1728 

\bibitem[Fontaine et al.(1985)]{Fontaineet85} Fontaine, G., 
Wesemael, F., Bergeron, P., Lacombe, P., Lamontagne, R., \& Saumon, D.\ 
1985, \apj, 294, 339                                                                                                   
\bibitem[Fontaine et al.(2001)]{Fontaineet01} Fontaine, G.,
Bergeron, P., Brassard, P., Bill{\` e}res, M., \& Charpinet, S.\ 2001,
\apj, 557, 792

\bibitem[Fontaine et al.(2003)]{Fontaineet03} Fontaine, G.,
Bergeron, P., Bill{\`e}res, M., \& Charpinet, S.\ 2003, \apj, 591, 1184
                                                                                                   
\bibitem[Gianninas et al.(2005)]{Gianninaset05} Gianninas, A., 
Bergeron, P., \& Fontaine, G.\ 2005, \apj, 631, 1100 

\bibitem[Giovannini et al.(1998)]{Giovanniniet98} Giovannini, O., 
Kepler, S.~O., Kanaan, A., Costa, A.~F.~M., \& Koester, D.\ 1998, \aap, 
329, L13 

\bibitem[Handler et al.(2002)]{Handleret02} Handler, G., 
Romero-Colmenero, E., \& Montgomery, M.~H.\ 2002, \mnras, 335, 399 

\bibitem[Hansen et al.(1985)]{Hansenet85} Hansen, C.~J., Winget, 
D.~E., \& Kawaler, S.~D.\ 1985, \apj, 297, 544 
 
\bibitem[Hesser et al.(1976)Hesser, Lasker, \& Neupert]{HesserLaskeraNeupert76} Hesser, J.~E., Lasker, 
B.~M., \& Neupert, H.~E.\ 1976, \apj, 209, 853 

\bibitem[Kanaan et al.(2002)Kanaan, Kepler, \& Winget]{Kanaanet02} Kanaan, A., Kepler, 
S.~O., \& Winget, D.~E.\ 2002, \aap, 389, 896 
                                                                                                   
\bibitem[Kanaan et al.(2005)]{Kanaanet05} Kanaan, A., et al.\ 
2005, \aap, 432, 219 

\bibitem[Kepler et al.(1982)]{Kepleret82} Kepler, S.~O., Nather, 
R.~E., McGraw, J.~T., \& Robinson, E.~L.\ 1982, \apj, 254, 676 

\bibitem[Kepler(1984)]{Kepler84} Kepler, S.~O.\ 1984, \apj, 286, 
314 

\bibitem[Kepler et al.(1995)]{Kepleret95} Kepler, S.~O., et al.\ 
1995, \apj, 447, 874 

\bibitem[Kepler et al.(2005)]{2005A&A...442..629K} Kepler, S.~O., 
Castanheira, B.~G., Saraiva, M.~F.~O., Nitta, A., Kleinman, S.~J., 
Mullally, F., Winget, D.~E., \& Eisenstein, D.~J.\ 2005, \aap, 442, 629 
 
\bibitem[Kim et al.(2005)]{Kimet05} Kim, A.\ et al. 2005,
astro-ph/0510104

\bibitem[Kleinman(1995)]{Kleinman95} Kleinman, S.~J.\ 1995, 
Ph.D.~Thesis, University of Texas at Austin

\bibitem[Kleinman et al.(1998)]{Kleinmanet98} Kleinman, S.\ J.\ et
al.\ 1998, \apj, 495, 424
                                                                                                   
\bibitem[Kleinman et al.(2004)]{Kleinmanet04} Kleinman, S.~J., et 
al.\ 2004, \apj, 607, 426 

\bibitem[Koester \& Vauclair(1997)]{KoesteraVauclair97} Koester, D., \& 
Vauclair, G.\ 1997, ASSL Vol.~214: White dwarfs, 429 
 
\bibitem[Koester \& Allard(2000)]{KoesteraAllard00} Koester, D.~\&
Allard, N.~F.\ 2000, Baltic Astronomy, 9, 119

\bibitem[Koester \& Holberg(2001)]{KoesteraHolberg01} Koester, D., \& 
Holberg, J.~B.\ 2001, ASP Conf.~Ser.~226: 12th European Workshop on White 
Dwarfs, 226, 299 
 
\bibitem[Kotak et al.(2002)Kotak, Kerkwijk, \& Clemens]{KotakKerkwijkaClemens02} Kotak, R., van Kerkwijk, 
M.~H., \& Clemens, J.~C.\ 2002, \aap, 388, 219 

\bibitem[McGraw \& Robinson(1975)]{McGrawaRobinson75} McGraw, J.~T., \& 
Robinson, E.~L.\ 1975, \apjl, 200, L89 

\bibitem[McGraw \& Robinson(1976)]{McGrawaRobinson76} McGraw, J.~T., \& 
Robinson, E.~L.\ 1976, \apjl, 205, L155 

\bibitem[McGraw(1979)]{McGraw79} McGraw, J.~T.\ 1979, \apj, 229,
203
                                                                                                   
\bibitem[McGraw(1980)]{McGraw80} McGraw, J.~T.\ 1980, NASA.~
Goddard Space Flight Center  Current Probl.~in Stellar Pulsation
Instabilities  p 501-512 (SEE N80-25229 15-90), 501

\bibitem[McGraw et al.(1981)]{McGrawet81} McGraw, J.~T., Fontaine, 
G., Lacombe, P., Dearborn, D.~S.~P., Gustafson, J., \& Starrfield, S.~G.\ 
1981, \apj, 250, 349                                                                                                    
\bibitem[Montgomery(2005)]{2005ApJ...633.1142M} Montgomery, M.~H.\ 2005, 
\apj, 633, 1142 
 
\bibitem[Mukadam et al.(2002)]{Mukadamet02} Mukadam, A.~S., Kepler, 
S.~O., Winget, D.~E., \& Bergeron, P.\ 2002, \apj, 580, 429 

\bibitem[Mukadam et al.(2003)]{Mukadamet03} Mukadam, A.~S.~et al.\
2003, \apj, 594, 961

\bibitem[Mukadam et al.(2004a)]{Mukadamet04a} Mukadam, A.~S., et al.\ 
2004a, \apj, 607, 982 

\bibitem[Mukadam et al.(2004b)]{Mukadamet04b} Mukadam, A.~S., Winget, 
D.~E., von Hippel, T., Montgomery, M.~H., Kepler, S.~O., \& Costa, 
A.~F.~M.\ 2004b, \apj, 612, 1052 

\bibitem[Mullally et al.(2005)]{Mullallyet05} Mullally, F., 
Thompson, S.~E., Castanheira, B.~G., Winget, D.~E., Kepler, S.~O., 
Eisenstein, D.~J., Kleinman, S.~J., \& Nitta, A.\ 2005, \apj, 625, 966 
 
\bibitem[Nather \& Mukadam(2004)]{NatheraMukadam04} Nather, R.~E.~\&
Mukadam, A.~S.\ 2004, \apj, 605, 846

\bibitem[Odonoghue \& Warner(1982)]{ODonoghueaWarner82} O'Donoghue, D.~E., 
\& Warner, B.\ 1982, \mnras, 200, 563 

\bibitem[O'Donoghue et al.(1992)O'Donoghue, Warner, \& Cropper]{ODonoghueWarneraCropper92} O'Donoghue, D., 
Warner, B., \& Cropper, M.\ 1992, \mnras, 258, 415 

\bibitem[Pesnell(1985)]{Pesnell85} Pesnell, W.~D.\ 1985, \apj,
292, 238

\bibitem[Pfeiffer et al.(1995)]{Pfeifferet95} Pfeiffer, B., et al.\ 
1995, Baltic Astronomy, 4, 245 

\bibitem[Robinson \& MacGraw(1976)]{RobinsonaMcGraw76} Robinson, E.~L., 
\& MacGraw, J.~T.\ 1976, \apjl, 207, L37 

\bibitem[Robinson(1979)]{Robinson79} Robinson, E.~L.\ 1979, IAU
Colloq.~53: White Dwarfs and Variable Degenerate Stars, 343

\bibitem[Robinson(1980)]{Robinson80} Robinson, E.~L.\ 1980, NASA.~ 
Goddard Space Flight Center  Current Probl.~in Stellar Pulsation 
Instabilities  p 423-451 (SEE N80-25229 15-90), 423 

\bibitem[Robinson et al.(1982)Robinson, Kepler, \& Nather]{Robinsonet82} Robinson,
E.~L., Kepler, S.~O., \& Nather, R.~E.\ 1982, \apj, 259, 219
                                                                                                   
\bibitem[Robinson et al.(1995)]{Robinsonet95} Robinson, E.~L.~et
al.\ 1995, \apj, 438, 908

\bibitem[Silvotti et al.(2005)]{2005A&A...443..195S} Silvotti, R., Voss, 
B., Bruni, I., Koester, D., Reimers, D., Napiwotzki, R., \& Homeier, D.\ 
2005, \aap, 443, 195 

\bibitem[Silvotti et al.(2005b)]{Silvottiet05b} Silvotti, R.\ \etal
2005b, \aap, submitted 

\bibitem[Stobie et al.(1993)]{Stobieet93} Stobie, R.~S., Chen, A., 
O'Donoghue, D., \& Kilkenny, D.\ 1993, \mnras, 263, L13 

\bibitem[Stobie et al.(1995)]{Stobieet95} Stobie, R.~S., 
O'Donoghue, D., Ashley, R., Koen, C., Chen, A., \& Kilkenny, D.\ 1995, 
\mnras, 272, L21 

\bibitem[Thompson et al.(2004)]{Thompsonet04} Thompson, S.~E., 
Clemens, J.~C., van Kerkwijk, M.~H., O'Brien, M.~S., \& Koester, D.\ 2004, 
\apj, 610, 1001 

\bibitem[Unno et al.(1989)]{Unnoet89} Unno, W., Osaki, Y., Ando, 
H., Saio, H., \& Shibahashi, H.\ 1989, Nonradial oscillations of stars, 
Tokyo: University of Tokyo Press, 1989, 2nd ed.

\bibitem[Vauclair et al.(1987)Vauclair, Dolez, \& Chevreton]{VauclairDolezaChevreton87} Vauclair, G., Dolez, 
N., \& Chevreton, M.\ 1987, \aap, 175, L13 

\bibitem[Vauclair et al.(1989)]{Vauclairet89} Vauclair, G., Goupil, 
M.~J., Baglin, A., Auvergne, M.~., \& Chevreton, M.\ 1989, \aap, 215, L17 

\bibitem[Vauclair et al.(1997)]{Vauclairet97} Vauclair, G., Dolez, 
N., Fu, J.~N., \& Chevreton, M.\ 1997, \aap, 322, 155 

\bibitem[Vauclair et al.(2000)]{Vauclairet00} Vauclair, G., Dolez, 
N., Fu, J.-N., Homeier, D., Roques, S., Chevreton, M., \& Koester, D.\ 
2000, \aap, 355, 291 

\bibitem[Weidemann (1990)]{Weidemann90} Weidemann, V.  1990, \araa,
28, 103

\bibitem[Winget(1982)]{Winget82} Winget, D.~E.\ 1982, 
Ph.D.~Thesis, University of Rochester

\bibitem[Winget \& Fontaine(1982)]{WingetaFontaine82} Winget, D.~E.~\&
Fontaine, G.\ 1982, Pulsations in Classical and Cataclysmic Variable Stars,
46

\bibitem[Wu(2001)]{Wu01} Wu, Y.\ 2001, \mnras, 323, 248 

\end{thebibliography}
\end{document}